\documentclass[twocolumn,showpacs,preprintnumbers]{revtex4}
\usepackage{graphicx}
\usepackage{dcolumn}
\usepackage{bm}

\begin{document}

\title{Comment on "Spin-Glass Solution of the Double-Exchange Model in
Infinite Dimensions"}
\author{Eugene Kogan$^{1}$ and Mark Auslender$^2$}
\affiliation{$^1$ Jack and Pearl Resnick Institute of Advanced Technology, Department of
Physics, Bar-Ilan University, Ramat-Gan 52900, Israel\\
$^2$ Department of Electrical and Computer Engineering, Ben-Gurion
University of the Negev, P.O.B. 653, Beer-Sheva, 84105 Israel}
\date{\today}

\begin{abstract}
\end{abstract}
 \maketitle

In a very interesting paper submitted to the Arxive some time ago
\cite{fishman}, R. S. Fishman, F. Popescu, G. Alvarez, T. Maier and
J. Moreno considered double-exchange model on a Bethe lattice in
infinite dimensions using dynamical mean-field theory. They analyzed
instabilities of the paramagnetic (PM) state with respect to
infinitesimal short-range magnetic order: for any site, a fraction
$1-q$ of the neighboring spins are the same and $q$ are opposite.
(In limiting cases $ q=0$ and $q=1$ the order is ferromagnetic and
antiferromagnetic respectively.) In this, mostly pedagogical note,
we present a somewhat different derivation of their general formula
for the magnetic disorder - order transition temperature and comment
on the interpretation of the intermediate phase obtained by the
authors.

The DE model, containing classical core spins and the conduction electrons
with the exchange coupling between them is a single electron one, and the
Hamiltonian can be presented as
\begin{equation}  \label{generic}
\hat{H}_{nn^{\prime}}=t_{n-n^{\prime}}-J \mathbf{m}_n\cdot
\hat{\mathbf{ \sigma}}\delta_{nn^{\prime}},
\end{equation}
where $t$ is the electron hopping, $J$ is the exchange coupling
$\hat{ \mathbf{\sigma}}$ is the vector of the Pauli matrices, and
$m_n$ is a vector of unit length which represent a core spin in a
classical model.

On the Bethe lattice (Caley tree) the local Green's function
\begin{equation}
\hat{g}(E)=(E-\hat{H})_{nn}^{-1}  \label{green}
\end{equation}
satisfies equation
\begin{equation}
\hat{g}_{n}=\frac{1}{E-\frac{W^{2}}{4}\hat{g}_{n+1}+J\mathbf{m}_{n}\cdot
\hat{\mathbf{\sigma }}},  \label{local}
\end{equation}
where $W$ is the bare (in the absence of exchange interaction) band width.
Further on all the energies we'll measure in the units of $W$.

In the framework of the DMFA \cite{georges} we substitute for Eq.
(\ref {local}) the following equation
\begin{equation}
\hat{G}_{n}=\left\langle \frac{1}{E-\hat{G}_{n+1}/4+J\mathbf{m}_{n}\cdot
\hat{\mathbf{\sigma }}}\right\rangle ,
\label{localCPA}
\end{equation}
where $G_{n}=\left\langle g_{n}\right\rangle $, $\left\langle
X(\mathbf{m} )\right\rangle \equiv \int
X(\mathbf{m})P(\mathbf{m})$, and $P(\mathbf{m})$ is a probability
of a given core-spin orientation (one-site probability). The
quantities $\hat{G}$ are $2\times 2$ matrices in spin space. In
the PM phase $P(\mathbf{m})=1$ (up to normalization), $G_{n}=G$,
and Eq. (\ref {localCPA}) is closed and takes the form
\begin{equation}
G=\frac{1}{2}\sum_{(\pm )}\frac{1}{E-G/4\pm J}.  \label{rq2}
\end{equation}

In the magnetically ordered phase the probability $P(\mathbf{m})$ should be
determined self-consistently with the solution of Eq. (\ref{localCPA}). The
DMFA approximation for the one-site probability $P(\mathbf{m})$ is:
\begin{equation}
P(\mathbf{m})\propto \exp \left[ -\beta \Delta \Omega (\mathbf{m})\right] ,
\label{prob2}
\end{equation}
where
\begin{equation}
\Delta \Omega (\mathbf{m})=\int_{-\infty }^{\mu }\Delta
D(E,\mathbf{m})dE, \label{probability44}
\end{equation}
and
\begin{equation}
\Delta D(E,\mathbf{m})=-\frac{1}{\pi }\mathrm{Im}\ln
\mathrm{det}\left[ E-
\hat{G}_{n+1}/4+J\mathbf{m}_{n}\hat{\mathbf{\sigma }}\right] ;
\label{probability4}
\end{equation}
the argument of both $G_{\mathrm{loc}}$ and $\Sigma $ is $E+i0$, and the
electron gas is considered as degenerate.

Eqs. (\ref{localCPA}) and (\ref{prob2}) present a complicated system of
equations. However, near the Curie temperature the system can be reduced to
an ordinary mean field equation. Eqs. (\ref{localCPA}) and (\ref{prob2}) can
be linearized with respect to small deviations of the locator from the
isotropic PM value \cite{kogan}. Looking for $G_{n}$ in the form
\begin{equation}
G_{n}=G+\mathbf{B}_{n}\hat{\mathbf{\sigma }},  \label{vectB-def}
\end{equation}
for the anisotropic part we obtain the following equation
\begin{equation}
(E-G/4)\mathbf{B}_{n}=\left( G/4+\frac{J^{2}G^{2}/6}{E-G/4}\right)
\mathbf{B} _{n+1}-GJ\mathbf{M}_{n},   \label{g}
\end{equation}
where $\mathbf{M}_{n}=\left\langle \mathbf{m} _{n}\right\rangle$.
Similarly, Eq. (\ref{prob2}) in linear approximation takes the form
\begin{equation}
P_{n}(\mathbf{m}_{n})\propto \exp \left( \mathbf{m}_{n}\cdot
\frac{\beta J}{2\pi }\int_{-\infty }^{\mu _{P}}\mbox{Im}\left[
\frac{G\mathbf{B}_{n+1}}{ E-G/4}\right] dE\right).
\label{probability}
\end{equation}

Ferromagnetic (FM) order is described by the equation
$\mathbf{B}_{n+1}= \mathbf{B}_{n}$
($\mathbf{M}_{n}=\mathbf{M}_{n+1}$) and antiferromagnetic (AFM)\
one by the equation $\mathbf{B}_{n+1}=-\mathbf{B}_{n}$
($\mathbf{M}_{n}=-\mathbf{M}_{n+1}$). Fishman et al. suggested a
more general ordering, using  some parameter $0\leq q\leq 1$,
described by
\begin{equation}
\mathbf{B}_{n+1}=(1-2q)\mathbf{B}_{n},
\label{mf3}
\end{equation}
so that $q=0$ corresponds to the FM and $q=1$ to the AFM order. After
substituting this prescription into Eq. (\ref{probability})  we obtain an
ordinary mean-field probability
\begin{equation}
P_{n}(\mathbf{m}_{n})\propto \exp \left[ -3\beta
T_{cr}(q)\mathbf{M}_{n}\cdot \mathbf{m}_{n}\right],
\label{prob-fin}
\end{equation}
where $T_{cr}(q)$ is a critical temperature given by
\begin{eqnarray}
&&T_{cr}(q) = (1-2q)\frac{J^{2}}{6\pi }\int_{-\infty }^{\mu
_{P}}dE \label{Theta}
\\
&&\mbox{Im}\left\{ \frac{G^{2}}{\left[ E-\frac{G}{4}\right] \left[
E-(1-q) \frac{G}{2}\right] -(1-2q)\frac{J^{2}G^{2}}{6}}\right\} ,
\nonumber
\end{eqnarray}
so that $T_{cr}\left( q=0\right) $ is the Curie and  $T_{cr}\left(
q=1\right) $ is the Neel temperature. Eq. (\ref{Theta}) is Eq. (6)
from the preprint \cite{fishman}, in which there was made the
transition from the summation with respect to discreet frequencies
to the integration with respect to energy.

Fishman et al. associated the values of $q\neq 0,1$ with a spin
glass phase. We think that  the phase taken into account by the
authors is a mixed state, combining both ferromagnetism and
antiferromagnetism, the shell $n$ and $n+1$ being the sublattices.
The order parameters are the vector of ferromagnetism (the averaged
magnetization) $(1-q)\mathbf{M}$ and vector of antiferromagnetism
(half of the difference between the magnetization of the
sublattices) $q\mathbf{M}$. From these definitions we immediately
obtain Eq. (\ref{mf3}).

\end{document}